# Formulation, Colloidal Characterization, and In Vitro Biological Effect of BMP-2 Loaded PLGA Nanoparticles for Bone Regeneration

**Teresa del Castillo-Santaella** [1], **Inmaculada Ortega-Oller** [2], **Miguel Padial-Molina** [2], **Francisco O'Valle**[3], **Pablo Galindo-Moreno** [2], **Ana Belén Jódar-Reyes** [1,4] **and José Manuel Peula-García** [1,5,*]

[1] Biocolloid and Fluid Physics Group, Department of Applied Physics, University of Granada, 18071 Granada, Spain
[2] Department of Oral Surgery and Implant Dentistry, University of Granada, 18071 Granada, Spain
[3] Department of Pathology, School of Medicine & IBIMER, University of Granada, 18071 Granada, Spain
[4] Excellence Research Unit "Modeling Nature" (MNat), University of Granada, 18071 Granada, Spain
[5] Department of Applied Physics II, University of Malaga, 29071 Malaga, Spain
* Correspondence: jmpeula@uma.es; Tel.: +34 952132722

**Abstract:** Nanoparticles (NPs) based on the polymer poly (lactide-co-glycolide) acid (PLGA) have been widely studied in developing delivery systems for drugs and therapeutic biomolecules, due to the biocompatible and biodegradable properties of the PLGA. In this work, a synthesis method for bone morphogenetic protein (BMP-2)-loaded PLGA NPs was developed and optimized, in order to carry out and control the release of BMP-2, based on the double-emulsion (water/oil/water, W/O/W) solvent evaporation technique. The polymeric surfactant Pluronic F68 was used in the synthesis procedure, as it is known to have an effect on the reduction of the size of the NPs, the enhancement of their stability, and the protection of the encapsulated biomolecule. Spherical solid polymeric NPs were synthesized, showing a reproducible multimodal size distribution, with diameters between 100 and 500 nm. This size range appears to allow the protein to act on the cell surface and at the cytoplasm level. The effect of carrying BMP-2 co-adsorbed with bovine serum albumin on the NP surface was analyzed. The colloidal properties of these systems (morphology by SEM, hydrodynamic size, electrophoretic mobility, temporal stability, protein encapsulation, and short-term release profile) were studied. The effect of both BMP2-loaded NPs on the proliferation, migration, and osteogenic differentiation of mesenchymal stromal cells from human alveolar bone (ABSC) was also analyzed in vitro.

## 1. Introduction

In the context of nanomedicine, tissue regeneration using colloidal micro- and nano-structures having unique size and surface activity has received increasing attention over recent years. Many efforts have been made to improve the engineering of these nano-systems in order to reach a "smart" delivery of bioactive molecules in order to optimize their therapeutic advantages and minimize harmful side effects [1]. With this aim, a broad spectrum of biocompatible nanocarriers has been described, showing properties suitable for different biological and therapeutic applications [2]. Among these varied proposals, polymeric nanosystems represent a major group in which poly lactic-co-glycolic acid (PLGA) is one of the most widely used due to its biocompatibility, biodegradability, and low cytotoxicity, gaining the approval from different drug agencies for human use [3,4].

PLGA-based structures are described as micro- and nanocarriers to deliver a wide variety of active molecules and drugs, synthetic or natural molecules with hydrophilic or hydrophobic properties, and biomolecules from proteins to nucleic acids [5–7]. PLGA micro- and nanosystems can be set up using different formulation techniques, with the possibility of a systemic or local distribution. These systems can be applied not only in tissue regeneration but also in very diverse therapies: Anticancer drug delivery, infections, inflammatory diseases, or gene therapy [3]. Despite this great potential, certain applications, especially in protein encapsulation, are hindered by problems, such as an uncontrolled release profile and protein denaturation [8–11].

The water-in oil-in water (W/O/W) double emulsion method is an "emulsion solvent evaporation" technique frequently used to encapsulate hydrophilic molecules as proteins in PLGA NPs [6,12]. The appropriate choice of organic solvents, the use of polymer-surfactant blends, and the addition of stabilizer-protective agents have proved to be key aspects for optimizing the resulting systems [9,11]. Additionally, a surface specific functionalization can be used to improve their versatility, allowing the chemical surface immobilization of different molecules in order to confer targeting or adhesive properties to these nanocarriers [13].

Within tissue engineering, bone regeneration has a broad range of applications, mostly in the field of dentistry, where PLGA is suggested as a reference polymer to formulate NPs with bone-healing uses [14]. The literature describes the delivery of bioactive molecules, normally growth factors, using polymeric microparticles (MPs) and NPs with PLGA as the main component [13]. Among the bone morphogenetic growth factors, BMP-2 (bone morphogenetic protein 2) has been the most frequently cited, with many examples in which encapsulation or surface adsorption enables adequate entrapment efficiency and diverse release patterns [15–19]. For proteins with a very short half-life, such as BMPs, biodegradable PLGA nanosystems provide protection and optimal dosage for an adequate stimulation of cell differentiation [20,21].

Thus, within this scenario, in the present work, we seek to optimize a nano-particulate system in order to carry out and control the release of BMP-2 using as a starting point the synthesis procedure of a lysozyme-loaded NP system, previously described for the encapsulation of that model protein [11]. Also, to encapsulate BMP-2, we prepared a second system in which this protein was co-adsorbed with bovine serum albumin onto the surface of empty NPs. The size and morphology, the protein encapsulation efficiency, the surface characteristics, and the colloidal and temporal stability were studied to complete the physico-chemical characterization of both NP systems.

The release profile of BMP-2 indicates the potential of a PLGA nanocarrier for bone regeneration and depends heavily on the polymer degradation by hydrolysis [22]. However, over the short term, during which the release does not depend on this chemical degradation, proper control of release is necessary in order to modulate other physical processes. Thus, we focused our release experiments on the short-term using different techniques to compare the two NP samples and establish the corresponding BMP-2 release profiles. Finally, the biological activity (cell migration, proliferation, and osteogenic differentiation) was tested in vitro using mesenchymal stromal cells (MSCs) derived from alveolar bone [23].

## 2. Materials and Methods

### 2.1. Nanoparticle Synthesis

2.1.1. Formulation

Poly(lactide-co-glycolide) acid (PLGA 50:50) ($[C_2H_2O_2]_x[C_3H_4O_2]_y$), x = 50, y = 50 (Resomer® 503H, (Evonik, Essen, Germany), 32–44 kDa was used as the polymer, and polymeric surfactant Pluronic F68 (Poloxamer 188) (Sigma-Aldrich, St. Louis, MO, USA) as the emulsifier. Their structure, based on a poly(ethylene oxide)-block-poly(propylene oxide)-block-poly(ethylene oxide), is expressed as $PEO_a$-$PPO_b$-$PEO_a$ with a = 75 and b = 30. Human recombinant bone morphogenetic protein, rhBMP-2 (Sigma-H4791), was used as therapeutic biomolecule. Water was purified in a Milli-Q Academic Millipore system. A double-emulsion synthesis method was used following a procedure previously described with slight modifications [11]. In this method, 100 mg of PLGA and 3 mg of deoxycholic acid (DC) were dissolved in a tube containing 1 mL of ethyl acetate (EA) and vortexed. In total, 40 µL of a buffered solution at pH 12.8, with or without rhBMP-2 (200 µg/mL), were added and immediately sonicated (Branson Ultrasonics 450 Analog Sonifier) for 1 min (Duty cycle dial: 20%, Output control dial: 4) with the tube surrounded by ice. This primary W/O emulsion was poured into a plastic tube containing 2 mL of a buffered solution (pH 12) of F68 at 1 mg/mL, and vortexing for 30 s. Then, the tube surrounded by ice was sonicated at the maximum amplitude for the micro tip for 1 min (Output control: 7). This second W/O/W emulsion was poured into a glass containing 10 mL of the buffered F68 solution and kept under magnetic stirring for 2 min. The organic solvent was then rapidly extracted by evaporation under vacuum to a final volume of 8 mL. The resulting empty and BMP-2 encapsulated NP systems were named NP and NP-BMP2, respectively. A detailed scheme of the synthesis procedure, with a yield based on the PLGA component always higher than 85%, is shown in Figure S1 of the Supplementary Materials.

2.1.2. Cleaning and Storage

After the organic solvent evaporation, the sample was centrifuged for 10 min at 20 °C at 12,000 rpm. The supernatant was filtered using Millipore nanofilters, 0.1 µm for measuring the free non-encapsulated protein. The pellet was then resuspended in phosphate buffer (1.15 mM $NaH_2PO_4$), PB, to a final volume of 4 mL and kept refrigerated at 4 °C. Under these conditions, the systems kept colloidal stability at least for one month.

2.1.3. Protein Loading and Encapsulation Efficiency

The initial protein loading was optimized for the nanoparticle formulation, preserving the final colloidal stability after the evaporation step and taking into account the amounts shown in the literature for this growth factor when encapsulated inside PLGA NPs [24,25]. Thus, we chose 2 µg as the initial total mass of rhBMP-2, which means a relation of $2 \times 10^{-5}\%$ *w/w* (rhBMP-2/PLGA). The amount of encapsulated rhBMP-2 was calculated by measuring the difference between the initial added amount, and the free non-encapsulated protein present in the supernatant after the cleaning step, which was tested by a specific enzyme-linked immuno-sorbent assay following the instructions of the manufacturer (ELISA, kit RAB0028 from Sigma-Aldrich, St. Louis, MO, USA). Then, protein-encapsulation efficiency (EE) was calculated as follows:

$$EE = \frac{M_I - M_F}{M_I} \times 100$$

where $M_I$ is the initial total mass of rhBMP-2, and $M_F$ is the total mass of rhBMP-2 in the aqueous supernatant.

2.1.4. Physical Protein Adsorption

Bovine serum albumin (BSA) and rhBMP-2 were coupled on the empty nanoparticle surface by a physical adsorption method. The appropriate volume of an aqueous protein solution containing 0.5 mg of BSA and 2 µg of rhBMP-2 was mixed with 5 mL of acetate buffer (pH 5) containing empty NPs with 12.5 mg of PLGA. This provided a starting amount of proteins corresponding to 0.04% *w/w* (protein/PLGA), while the mass relation between proteins was 0.4 *w/w* (rhBMP-2/BSA). This solution was incubated at room temperature for 2 h under mechanical stirring. The nanoparticles were separated from the buffer solution by centrifugation, and after the supernatants were filtered (Millipore nanofilters, 0.1 µm), they were qualitatively analyzed by gel electrophoresis while the protein quantification was made by a bicinchoninic acid protein assay (BCA) (Sigma-Aldrich, St. Louis, MO, USA) for BSA and the specific ELISA for rhBMP-2. The nanoparticle pellet was resuspended in phosphate buffer (pH 7.4) and stored at 4 °C. This system was named NP-BSA-BMP2.

2.1.5. Protein Separation by Gel Electrophoresis, SDS-PAGE

The protein-loaded NPs and different supernatants were treated at 90 °C for 10 min in the following buffer: 62.5 mM Tris-HCl (pH 6.8 at 25 °C), 2% (*w/v*) sodium dodecyl sulfate (SDS), 10% glycerol, 0.01% (*w/v*) bromophenol blue, 40 mM dithiothreitol (DTT). Samples were then separated by size in porous 12% polyacrylamide gel (1D SDS polyacrylamide gel electrophoresis), under the effect of an electric field. The electrophoresis was run under constant voltage (130 V, 45 min) and the gels were stained using a Coomassie Blue solution (0.1% Coomassie Brilliant Blue R-250, 50% methanol and 10% glacial acetic acid) and destained with the same solution lacking the dye.

*2.2. Nanoparticle Characterization: Morphology, Size, Concentration, and Electrokinetic Mobility*

NPs were imaged by scanning electron microscopy (SEM) with a Zeiss SUPRA 40VP field-emission scanning electron microscope from the Scientific Instrumentation Center of the University of Granada (CIC, UGR).

The hydrodynamic size distribution of the NPs was evaluated by nanoparticle tracking analysis (NTA) with a NanoSight LM10-HS (GB) FT14 (NanoSight, Amesbury, UK) and an sCMOS camera. The particle concentration according to the diameter (size distribution) was calculated as an average of at least three independent size distributions. The total concentration of NPs of each system was determined in order to control the number of particles used in cell experiments. The measurement conditions for all samples were 25 °C, a viscosity of 0.89 cP, a measurement time of 60 s, and a camera gain of 250. The camera shutter was 11 and 15 ms for the empty and BMP-loaded NPs, respectively. The detection threshold was fixed at 5.

The electrophoretic mobility of the NPs was determined using a Zetasizer® NanoZeta ZS device (Malvern Instrument Ltd., Malvern, UK) working at 25 °C with an He-Ne laser of 633 nm, and a 173° scattering angle. Each data point was taken as an average over three independent sample measurements. For each sample, the electrophoretic mobility distribution and the average electrophoretic mobility (μ-average) were determined by the technique of laser Doppler electrophoresis.

*2.3. Colloidal and Temporal Stability in Biological Media*

The average hydrodynamic diameter and the polydispersity index (PDI) by dynamic light scattering (DLS) of each NP system were measured in different media (phosphate buffer (PB) saline phosphate buffer (PBS), and cell culture medium: Dulbecco's modified Eagle's medium, DMEM (Sigma)). Also, data on temporal stability were gathered by repeating these analyses at different times after synthesis (0, 1, and 5 days) and after 1 month under storage conditions.

In vitro release experiments were conducted as follows: 1 mL of each sample for each incubation time was suspended in PBS at 37 °C. After the corresponding time (24, 48, 96, 168 h), NPs were separated from the supernatant of released proteins by centrifugation for 10 min at 14,000 rpm (10 °C). The NP pellet was suspended in 1 mL of 0.05 M NaOH and stirred for 2 h for a complete polymer degradation. The alkaline protein solution was assayed by BCA and ELISA to quantify the unreleased amount. The protein released was calculated taking into account the total encapsulated amount. All experiments were made in triplicate.

*2.4. Cell Interactions*

For all biological in vitro studies, a cell population cultured from the maxillary alveolar bone was used. This population was previously characterized and confirmed to present all characteristics of a mesenchymal stromal cell population (MSC) [23]. Cells were taken from healthy human donors after the approval from the Ethics Committee for Human Research from the University of Granada (424/CEIH/2018). Regular Dulbecco's modified Eagle's medium (DMEM) with 1 g/L glucose (DMEM-LG) (Gibco), 10% fetal bovine serum (FBS) (Sigma-Aldrich, St. Louis, MO, USA), 1:100 of non-essential amino acid solution (NEAA) (Gibco), 0.01 μg/mL of basic fibroblast growth factor (bFGF) (PeproTech, London, UK), 100 U/mL of penicillin/streptomycin, and 0.25 μg/mL of amphotericin B was used as culture medium for all experiments. Cultures were maintained at 37 °C in a 5% $CO_2$ atmosphere (2000 cells/well). All biological experiments were repeated in triplicate at least 3 times per condition.

2.4.1. Cell Migration

A cell-migration assay was conducted as previously described [26,27]. Briefly, MSCs were distributed on to three wells for each condition and allowed to grow to a cell confluency close to 99%, in 24-wells/plate at 3000 cells/cm$^2$, and in each well three different scratches were made. Then, cells were starved for 24 h by adding culture medium without serum. A scratch was made using a pipette tip along the diameter of the well. A wash step with PBS was performed to remove the scratched cells. Fresh complete culture media was added and supplemented depending on the assigned group (BMP-2, NP-BMP2, and NP-BSA-BMP2 at 1.25, 2.5, and 5 ng/mL of BMP-2). Afterwards, nine images were taken from the same area in each condition until 48 h later. On these images, the scraped area was measured by ImageJ software (National Institute of Health, Bethesda, MD, USA; http://rsbweb.nih.gov/ij/). The reduction in the scratched area over time was measured considering the area at time 0 as 100% open.

2.4.2. Cell Proliferation

Proliferation was evaluated by a sulphorhodamine (SRB) assay [28]. The assay was conducted by seeding the cells at 1500 cells/cm$^2$ in a 96-well plate at a confluence not higher than 50%. After cell attachment, the different supplements were added (BMP-2, NP-BMP2, and NP-BSA-BMP2 at 1.25, 2.5, and 5 ng/mL of BMP-2) and the cells were maintained in culture for up to 7 days. At each time point, the cells were washed with 1X PBS and fixed by adding ice-cold 10% trichloroacetic acid for 20 min at 4 °C. Then, the cells were washed 3 times with $dH_2O$ and dried until all time points were collected. Each well received 0.4% SRB in 1% acetic acid for 20 min at room temperature with gentle shaking. The staining was finished by washing each well 3 times with 1% acetic acid and drying it at room temperature for 24 h. The dye was retrieved from the cells by adding 10 mM Tris Base at pH 10.5 and gently shaking for 10 min. The solution recovered was then distributed in a 96-well plate and the optical absorbance was read at 492 nm.

2.4.3. Osteogenic Differentiation

Osteogenic differentiation was evaluated by adding osteogenic media to the cell culture in combination with free BMP-2, NP-BMP2, and NP-BSA-BMP2 at the highest dosages used in previous experiments. Cells were seeded at 3000 cells/cm$^2$ and cultured to reach an 85% to 90% confluency. This was followed by the addition of induction media containing 10 mM of β-glycerophosphate (Fluka, 50020), 0.1 μM of dexamethasone (Sigma-Aldrich, D2915) and 0.05 mM of L-ascorbic acid (Sigma-Aldrich, A8960). Cell cultures were maintained for 7 days to analyze early activity. At day 7, cells were collected in 1 mL of TRIzol®. Then, RNA was extracted and converted to cDNA. Alkaline phosphatase (ALP) was then evaluated, expression being calculated relative to glyceraldehyde-3-phosphate dehydrogenase protein (GAPDH) by the $2^{-\Delta\Delta Ct}$ method. These procedures were conducted as described elsewhere [23]. Forward and reverse primer sequences were AGCTCATTTCCTGGTATGACAAC and TTACTCCTTGGAGGCCATGTG for GAPDH, and TCCAGGGATAAAGCAGGTCTTG and CTTTCTCTTTCTCTGGCACTAAGG for ALP.

2.4.4. Statistical Evaluation

Cell migration and proliferation were evaluated by ANOVA followed by Tukey multiple comparisons test for pairwise analysis. Comparison between the levels of ALP at 4 vs. 7 days were analyzed by paired Student's *t* test. In all cases, a *p* value lower than 0.05 was established as statistical significance.

## 3. Results and Discussion

*3.1. Nanoparticle Formulation*

Double emulsion-solvent evaporation has been described as a robust and frequently used method to produce biomolecule-loaded PLGA NPs [6,12,13,29]. A formulation previously optimized by our group enabled the preservation of the biological activity of encapsulated biomolecules using a slightly aggressive organic solvent. Moreover, deoxycholic acid has been used in the first step of the formulation in order to improve the colloidal stability of NPs and, simultaneously, to obtain NP surfaces enriched with carboxylic groups, improving their versatility and allowing a subsequent chemical immobilization of different specific ligands [30]. By means of this improved formulation, in the present work, we developed empty nanoparticles (NPs) or nanoparticles encapsulating rhBMP-2 (NP-BMP2). A schematic description of the synthesis procedure is shown in Figure S1 of the Supplementary Data. For NP-BMP2, we achieved a protein-encapsulation efficiency (EE) of 97 ± 2%. This result is consistent with the literature in which several authors have reported similarly high values encapsulating this protein inside PLGA nano- and microparticles [31,32]. Our formulation has several factors leading to this very high EE value: The low protein/polymer relation in mass [33], the affinity of rhBMP-2 to an unspecific interaction with hydrophobic surfaces [31], or the addition of stabilizers (poloxamer) in the second step of the double-emulsion procedure [13]. The absence of rhBMP-2 in the supernatant resulting from the centrifugation step in the cleaning process was verified by ELISA and SDS-PAGE, in which a clear band corresponding to 14 kD of rhBMP-2 polypeptidic chains is shown for lane A in Figure 1, corresponding to NP-BMP2. The mass of protein encapsulated, around 2 µg, is similar to that of different PLGA micro- and nanosystems described in the literature [18,34,35]. Taking into account the storage conditions for our samples, this corresponds to 500 ng/mL, which represents a sufficient concentration for practical applications since this growth factor shows in vitro biological activities at very low dosages (5–20 ng/mL) [13].

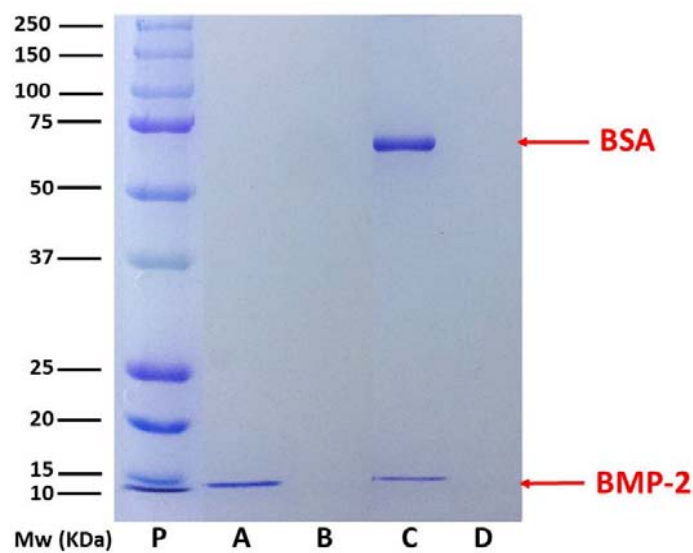

**Figure 1.** SDS polyacrylamide gel electrophoresis (SDS-PAGE) analysis under reducing conditions of solid PLGA Nanoparticles (PLGA NPs) and liquid (supernatant) fractions of different NP systems. Lane **P**: Protein standards; lane **A**: NP-BMP2 (bone morphogenetic protein); lane **B**: supernatant of NP-BMP2 after synthesis and encapsulation of rhBMP-2; lane **C**: NP after physical adsorption of BSA/rhBMP-2; lane **D**: supernatant after physical adsorption of BSA(bovine serum albumin)/rhBMP-2 on NP system.

On the other hand, a second nanosystem resulted, modifying the way in which rhBMP-2 is incorporated in the nanocarrier. There are several examples of surface adsorption of different growth factors in micro- and nanoparticles [35–37], and surface immobilization over the encapsulation has recently been proposed as a way to modulate the later release of biomolecules. This process, which depends on the slow diffusion of biomolecules through the polymeric matrix, is consequently highly influenced by the protein–polymer interaction [38,39] and polymer degradation [3,6]. Thus, this new focus on the use of PLGA NPs for biomolecule delivery was explored by immobilizing the protein rhBMP-2 on the surface of empty NPs by means of simple physical adsorption. This process is known to be governed by electrostatic and hydrophobic interactions between protein molecules and NP surfaces [40].

For this, the surface-charged groups, the hydrophilicity, the net charge of the protein molecules, and the characteristics of the adsorption medium are the reference parameters. Thus, we designed a co-adsorption experiment in which a mixture of rhBMP-2 and BSA (0.4% *w/w*, rhBMP-2/BSA) interact simultaneously with the PLGA NP surface. Albumins are routinely used as protective proteins when growth factors are incorporated in PLGA NPs [13,19]. Moreover, a surface distribution of BSA molecules can improve the colloidal stability of NPs at physiological pH due to their net negative charge under these conditions [41]. Figure S2 from Supplementary Materials shows a scheme of the co-adsorption process. The adsorption efficiency is higher than 95% and in SDS-PAGE from Figure 1, two bands characteristic of both proteins can be seen in lane C, corresponding to the NP-BSA-BMP2 nanosystem. However, lane D, corresponding to the run of the supernatant from the centrifugation of the nanosystem after adsorption processes, shows the absence of any protein. This result is fully explained by taking into account the pH of the medium (pH 5.0), near the isoelectric point of BSA, where the adsorption of this protein onto negatively charged nanoparticles presents a maximum [40,42]. The immobilization of rhBMP-2 on the negatively charged surface of NPs proves they are electrostatically favored due to the positive net charge of this protein at acid and neutral pH.

*3.2. Nanoparticle Characterization*

3.2.1. Nanoparticle Size

SEM and STEM micrographs (Figure 2) show that the samples consist of spherical particles of different diameters (between 150 and 450 nm), a range similar to that found in a previous work in which NPs were loaded with lysozyme following a similar synthesis protocol [11]. In that work, the DLS technique failed to provide a reliable size distribution. Therefore, the NTA technique was directly used to determine the hydrodynamic size of the BMP2-loaded NPs (see NTA video in the Supplementary Material).

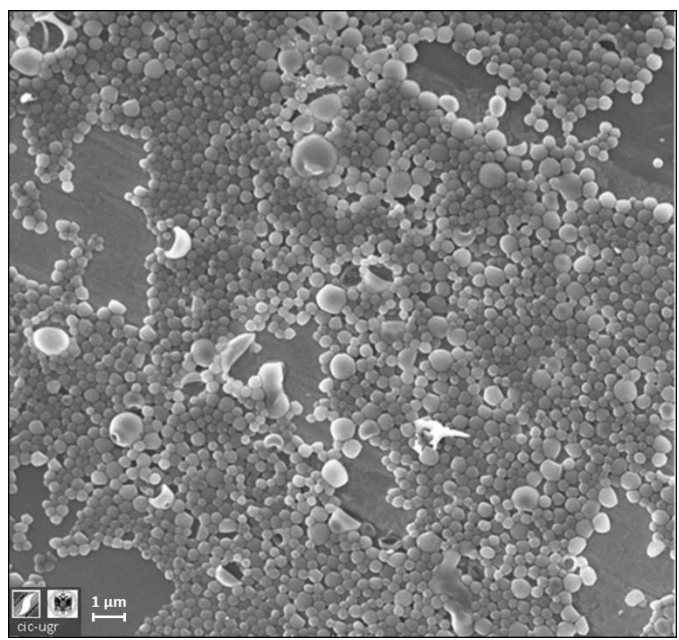

**Figure 2.** Scanning electron microscopy (SEM) micrograph of rhBMP-2-loaded nanoparticles (NP-BMP2).

The size distributions for empty (NP) and BMP-loaded NPs (NP-BMP2) from NTA (Figure 3 and videos S1, S2) were consistent with the SEM images. Particles with diameters between 100 and 500 nm were found to have the highest particle concentration at around 200 nm. The loading with BMP had an effect on the size distribution, leading to more defined peaks. These measurements enabled us to determine the concentration of particles in the measured sample: $6.88 \pm 0.09 \times 10^8$ pp/mL and $5.19 \pm 0.12 \times 10^8$ pp/mL for NP and NP-BMP2 nanosystems, respectively. These values were used (by taking into account the corresponding dilution) to control the number of particles added in the cell experiments.

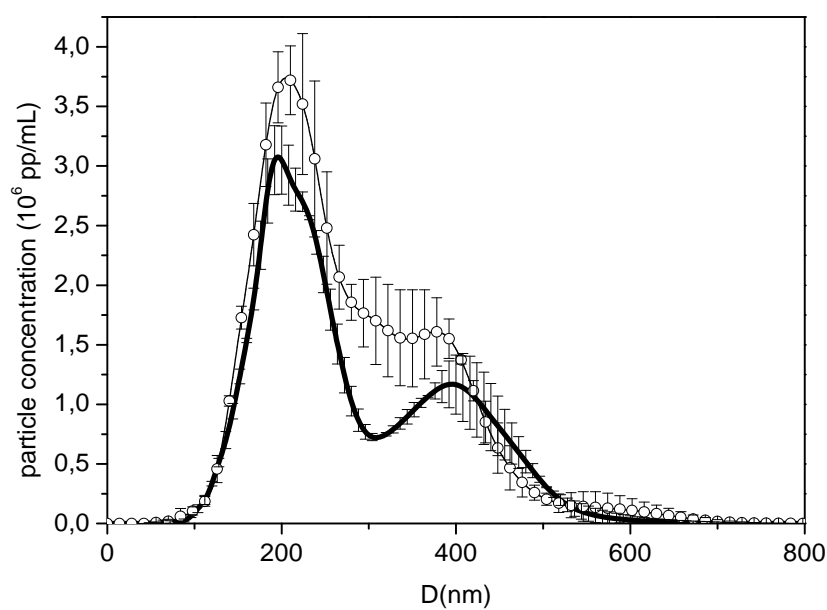

**Figure 3.** Hydrodynamic diameter distribution of NP (circles) and NP-BMP2 (thick black line) measured at pH 7.0 (phosphate buffer) by nanoparticle tracking analysis (NTA).

3.2.2. Electrokinetic Mobility and Colloidal Stability

The surface charge of nanoparticles can be analyzed using an electrokinetic study by measuring the electrophoretic mobility ($\mu_e$) under different conditions. Figure 4 shows the $\mu_e$ and zeta potential values for the three nanosystems: NP, NP-BMP2, and NP-BSA-BMP2, at low ionic strength and different pH values. The electric surface charge of NPs resides in the carboxylic groups of the uncapped PLGA and deoxycholic acid molecules. These functionalized groups are additionally useful due to the possibility of a chemical surface vectorization in order to develop directed delivery nanocarriers [43]. It was previously confirmed that protonation of these acidic surface groups at pH values under their pKa value was tightly correlated with a loss of surface charge and consequently a reduction (in absolute value) of the electrophoretic mobility of the colloidal system [44,45]. Usually, when colloidal particles are coated by protein molecules, the $\mu_e$ values change markedly compared with the same bare surfaces and are influenced by the electrical charge of the adsorbed protein molecules [46,47]. The electrokinetic behavior of the NP-BMP2 system remains similar to that of NP, and encapsulation of rhBMP-2 does not affect the surface charge distribution. A similar result was reported by d'Angelo et al. on encapsulating different growth factors in PLGA-poloxamer blend nanoparticles in the same proportion *w/w* of protein/polymer [24]. This may be due to the low amount of encapsulated protein and its distribution in the inner part of the NPs (far from the surface). In our system, this internal distribution may be favored by the encapsulating conditions where the basic pH (pH 12.0) of the water phase containing rhBMP-2 allows a negative charge of these protein molecules, thereby preventing their electrostatic specific interaction with acidic groups of the NPs.

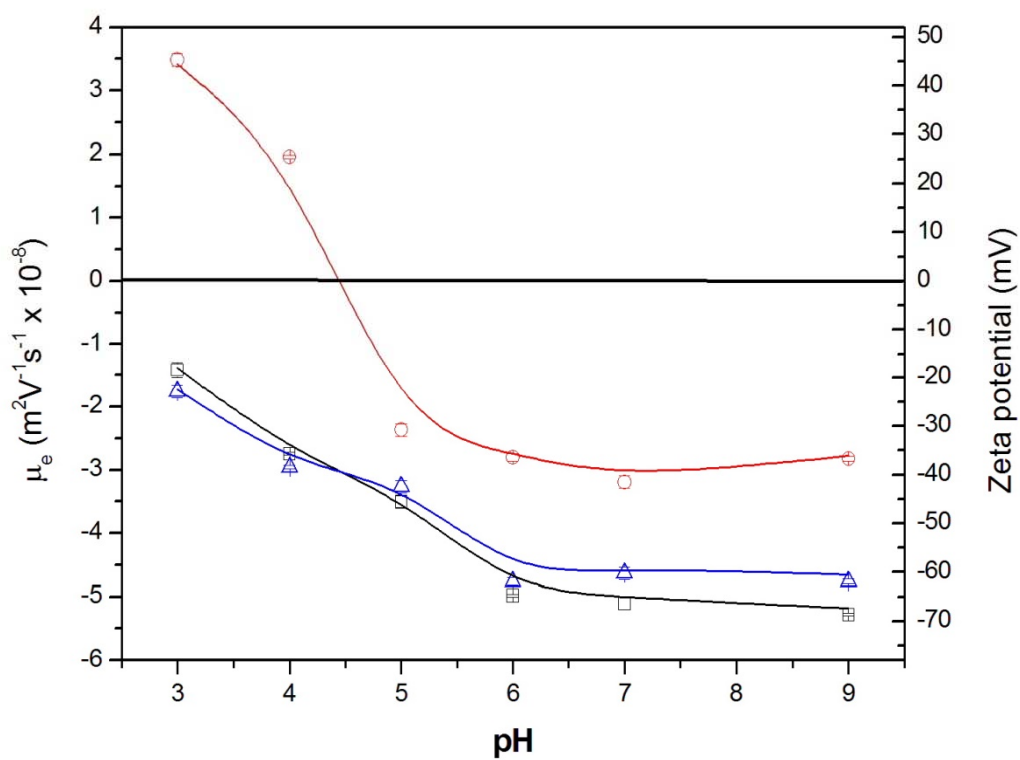

**Figure 4.** Electrophoretic mobility and zeta potential vs. pH in buffered media of low salinity (ionic strength equal to 0.002 M) for the different nanosystems: (black square) NP; (blue triangle) NP-BMP2; (red circle) NP-BSA-BMP2.

The electrokinetic distribution for the NP-BSA-BMP2 system radically changes. As previously shown, the very high adsorption efficiency leads to NPs with both proteins adsorbed around their surface. This situation is closely correlated with the $\mu_e$ values from Figure 4. Taking into account the *w/w* relation between adsorbed proteins (250 times higher for BSA), albumin molecules modulate the behavior at pH values below their isoelectric point (pI 4.7), where the positive net charge of BSA masks the original surface charge of NPs and even changes their original values to positive ones. This is a typical result found for this protein-covering colloidal particles [42,48]. At neutral and basic pH values, BSA molecules have a negative net charge, and the slight decrease in the absolute $\mu_e$ values could be due to the reduction of the negative net surface charge of NPs, which may be shielded, at least in a small part, by the positive charge of rhBMP-2 molecules under their basic isoelectric point (pI 9.0).

The colloidal stability for the different nanosystems (NP, NP-BMP2, and NP-BSA-BMP2) was determined by analyzing the size distributions in various media (PB, PBS, and DMEM) at different times after synthesis (0, 1, and 5 days). Size distributions similar to the original ones were found for the two formulations, NP and NP-BMP2, in all the media analyzed. This result was similar to that previously found for these types of NPs encapsulating lysozyme [11], in which the combination of electrostatic and steric interactions generated by surface chemical groups of NPs confer the stability mechanism that prevents colloidal aggregation [33]. The decrease of the absolute value of the zeta potential for the NP-BSA-BMP2 system as a consequence of surface protein distribution does not affect its colloidal stability. This system also maintains the same size distribution in the different media. It is commonly accepted that a zeta potential higher than +30 or −30 mV will give rise to a stable colloidal system [49] and the zeta potential value for NP-BSA-BMP2 is above −30 mV. Colloidal stability in PBS and DMEM, typically used media for the development of scaffold or cell interactions, respectively, assures the potential use of these nanosystems for in vitro or in vivo living environments. Additionally, these systems maintained their size under storage in PB, at 4 °C for at least 1 month (data not shown), showing this to be an adequate medium for sample storage.

### 3.2.3. Protein Release

One of the main problems for micro- or nanosystems of PLGA drug delivery is to find the appropriate release pattern for encapsulated/attached protein molecules. A wide spectrum of formulations modulates this property by the use of different types of synthesis processes, PLGA polymers, co-polymers, and stabilizers [3,13]. An adequate limitation and control in the burst release is critical for BMPs in order to ensure long-term continuous release that, favored by the polymer degradation, provides better in vivo action in driving bone and cartilage regeneration [20]. Therefore, we previously developed a dual PLGA nanosystem for controlled short-term release, where protein diffusion and protein–polymer interaction are the main factors governing this process [11].

In the present work, NP-BMP2 and NP-BSA-BMP2 nanosystems represent two different ways in which rhBMP-2 was incorporated into the nanocarrier. Figure 5A shows the cumulative release of both proteins, rhBMP-2 and BSA, for different systems as a function of time in a short-term period (7 days). The encapsulated rhBMP-2 protein reaches an amount released of around 30% of the initial encapsulated one while adsorbed rhBMP-2, despite its surface distribution, is three times lower. However, BSA shows released amounts up to 80% of the initial adsorbed ones. In all cases, error bars correspond to the standard deviations from three independent experiments. Under these conditions, the growth factor encapsulated in NP-BMP2 presents a release pattern similar to that previously found with the same formulation but using lysozyme as the protein [11]. Poloxamer in the water phase of the synthesis process can be key in modulating both specific and unspecific interfacial protein interactions [50]. Thus, the relation between protein–polymer interaction and protein diffusion appears to be well balanced, preventing an excessive initial burst and simultaneously maintaining the needed protein flux to release around a third of the encapsulated rhBMP-2 in 7 days. Although an excessive initial burst has been widely reported for PLGA NPs related with protein molecules close to the surface [6], this situation did not appear for the NP-BMP2 system, this being consistent with the electrokinetic behavior that did not show the presence of protein near surface. The literature offers some examples with reduced short-term release of BMP-2 using more hydrophilic PLGA-PEG co-polymers [16] or a different synthesis process [25].

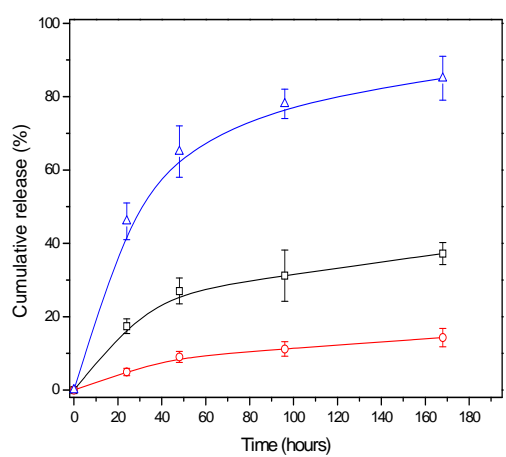 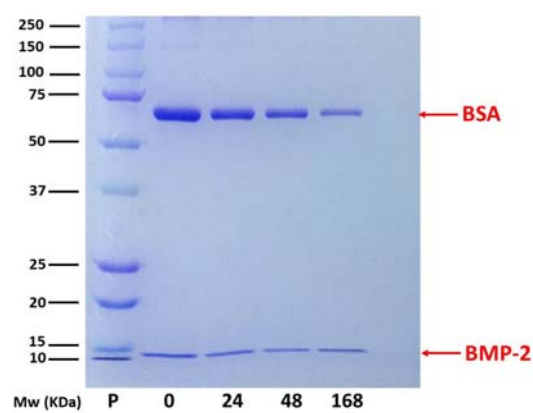

(A)              (B)

**Figure 5.** (**A**) Cumulative release of rhBMP-2 for NP-BMP2 (black square) and NP-BSA-BMP2 (red circle) systems; and cumulative release of BSA for NP-BSA-BMP2 (blue triangle) system, incubated for different times at 37 °C in saline phosphate buffer (pH 7.4). (**B**) SDS-PAGE analysis under reducing conditions of solid fraction of NP-BSA-BMP2 after release at different times where the number of each lane corresponds to the time in hours.

The release performance for the NP-BSA-BMP2 system, also shown in Figure 5A, presents notable differences. The electrokinetic profile has previously justified the surface location of BSA and rhBMP-2 on the surface, which could lead to a fast release of both proteins. However, results from Figure 5A,B show this trend only for the BSA protein that is released from NPs, with about 20% of the initial amount remaining after seven days. However, up to 90% of the initial load of rhBMP-2 protein, unlike BSA, remains attached to the surface. The NP surface with hydrophilic groups form poloxamer molecules and a negative charge due to the abundant presence of carboxylic groups (end-groups of PLGA and deoxycholic acid molecules) favor a desorption process for BSA, whose molecules have a negative charge under release conditions (physiological pH). This agrees with the results of other authors who, even after encapsulating BSA in PLGA-poloxamer blend NPs, achieved a fast burst release of above 40% to 50% of the initial protein amount [33]. Moreover, the co-encapsulation of albumins with growth factors could strongly affect its release profile, causing an initial burst [21,24]. Otherwise, the specific electrostatic attraction between positive rhBMP-2 molecules and negative surface groups slows down the short time release of this protein. This result is in agreement with the low release of adsorbed BMP previously found using PLGA micro- and nanoparticles with uncapped acid end groups [38,51]. Thus, the combination of different methods for trapping BMP-2 into and around NPs shows up the possibility of attaining a properly controlled release, balancing the interactions between polymers, stabilizers, and protein.

*3.3. Biological Activity and Interactions*

3.3.1. Cell Migration

Cell migration is the first and necessary step in tissue regeneration [52]. Thus, a regenerative agent must accelerate cell migration or, at least, not interfere with it. In the present study, we found no differences between the groups, doses, and control in terms of closure of a scratched area (ANOVA with Tukey multiple comparisons test) (Figure 6). In contrast to our findings, previously published data suggests a positive effect of BMP-2 on cell migration [53,54]. However, in those studies, the doses applied, and the cell types were different than in the current experiments. We used lower doses of BMP-2 in order to test whether, even at low dosages, BMP-2 could still provide benefits if protected in a nanoparticle system. As mentioned, we demonstrated no negative effect of the system on cell migration. Our results nonetheless support the idea that BMP-2 activity is mediated by the activation of the phosphoinositide 3-kinase (PI3K) pathway, a common group of signaling molecules that participate in several process with BMP-2 and other molecules [26,54]. It should also be mentioned that the timeframe of a migration assay is short. Thus, the potential advantages of a controlled-release system as the one under study might be limited. That is, the release of BMP-2 from the nanoparticles, as demonstrated in Figure 5, is limited to the first 48 h. Thus, a sustained positive effect on migration activity over time could be hypothesized.

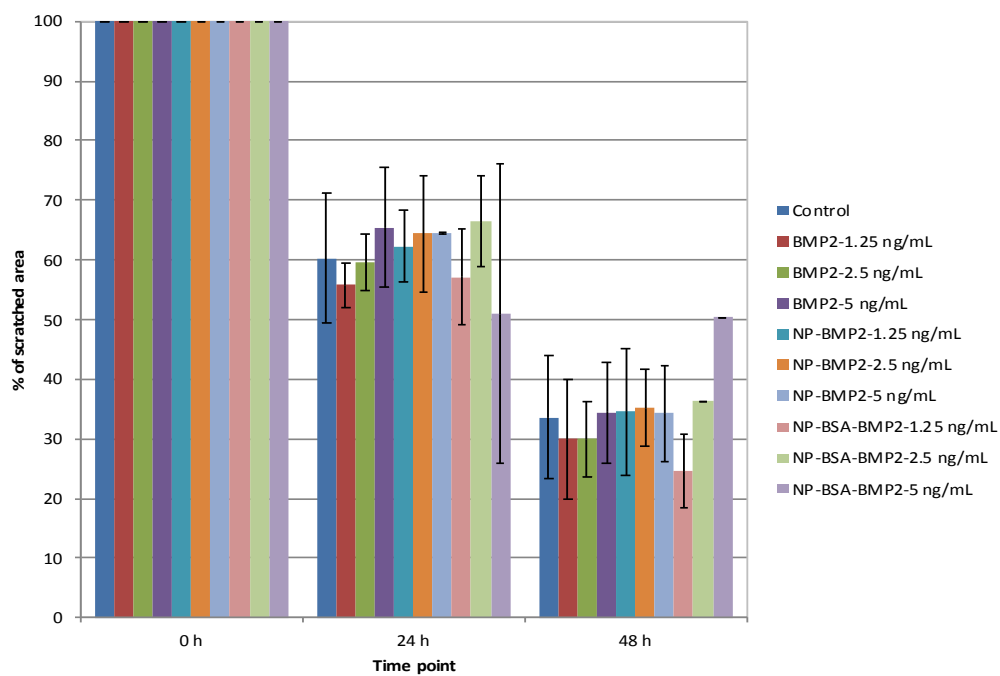

**Figure 6.** Migration assay. Percentage of scratched area closure at 24 and 48 h on different groups and doses.

3.3.2. Cell Proliferation

Proliferation is another of the cell activities required for tissue regeneration. However, this property must be balanced with both migration and differentiation, and not all three characteristics increase at the same time and with the same ratios [55]. In fact, reportedly, when a dose of BMP-2 induces higher proliferation, it decreases differentiation [56]. This property has been extensively analyzed but discrepancies can still be detected in the literature. Therefore, Kim et al. analyzed different doses of BMP-2 and its effect on cell proliferation and apoptosis. It was confirmed in vitro that high doses, but still lower than those used clinically, reduce cell proliferation and increase apoptosis [57]. This should be avoided. We have found that although free BMP-2 does not induce higher proliferation than the control at any of the doses applied nor time points (ANOVA with Tukey multiple comparisons test), the same amount of BMP-2 encapsulated or adsorbed onto PLGA nanoparticles boosts proliferation, this being statistically significant when using a dose of 2.5 ng/mL or higher (ANOVA with Tukey multiple comparisons test) (Figure 7). These dosages are still lower than those suggested in previous studies. Apart from that difference, a positive effect on proliferation was still achieved. Moreover, following the release pattern from Figure 5, more BMP-2 is expected to be released over time beyond the 7-day time frame. Thus, a sustained induction effect could be expected as well until full confluency of the cell culture.

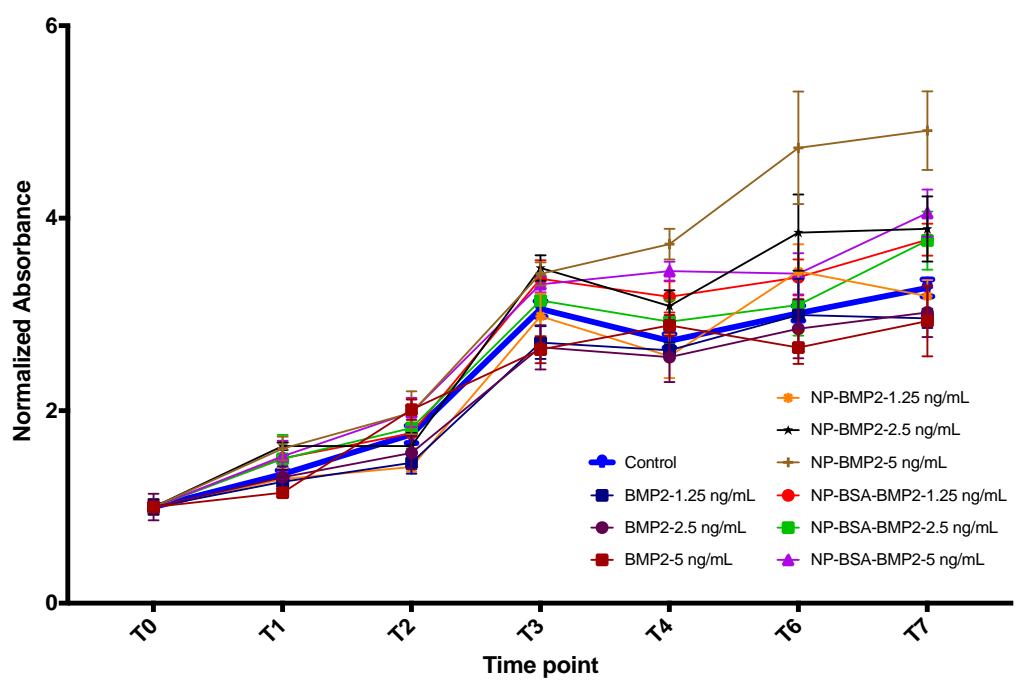

**Figure 7.** Proliferation of human mesenchymal stromal cells (MSCs) as measured by sulphorhodamine (SRB) absorbance. Results were normalized to T0 in each group.

3.3.3. Osteogenic Differentiation

It has been confirmed that cell differentiation induced by BMP-2 needs the presence of permissive osteoinductive components. Particularly, β-glycerophosphate has been shown to exert a synergistic effect with BMP-2 in inducing cell differentiation [56]. Thus, to test for osteogenic differentiation, we analyzed the expression of ALP mRNA. Maximum ALP activity was found to occur 10 days after stimulation with PLGA-based microparticles containing BMP-2 in co-encapsulation with human serum albumin [16]. Although other tests could have been used to reinforce our findings, ALP is known to modulate the deposition of mineralized nodules, thus indicating osteoblastic activity. For all of this, we supplemented the differentiation media with β-glycerophosphate and either free BMP-2, NP-BMP2, or NP-BSA-BMP2 for 4 and 7 days so that we could capture the early dynamics of the expression of the gene. In our study, we identified an increase in the expression of ALP in all groups from day 4 to day 7 (Figure 8). Although ALP at day 7 in the BMP-2 group appears to be higher than for the other two groups, the change did not prove significant. In fact, differences between groups were not statistically significant within any time period. Noteworthy though, the increase was not significant within the BMP-2 group ($p = 0.141$, Student's *t* test), but it was significant within the other two groups ($p = 0.025$ and $p = 0.003$; NP-BMP2 and NP-BSA-BMP2 groups, respectively). This, again, could be taken as a confirmation of the sustained release of the protein from the nanoparticle system beyond the earlier time points.

This and both the migration and proliferation studies described below lead us to confirm that the system proposed can maintain a proper release of BMP-2 over time, sustaining a positive effect on cell migration and proliferation with initial reduced doses of BMP-2. The fact that the excessive initial burst is prevented is important for the application of this nanotechnology in bone regeneration, as in dentistry. In this way, the negative effects of initial high doses of BMP-2 are avoided at the same time as the molecule is protected from denaturalization inside the NP. Thus, the regenerator effects are maintained over time.

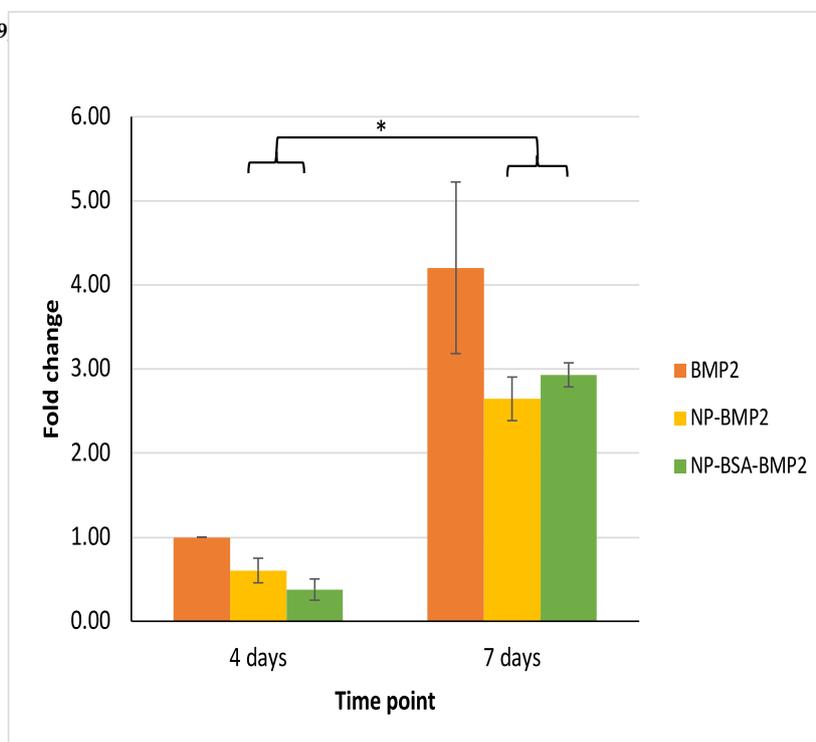

**Figure 8.** Relative fold change in the expression of ALP mRNA (control group: BMP2 at 4 days). * = Statistical significance of the comparison over time (*p* = 0.025 and *p* = 0.003, Student's *t* test; NP-BMP2 and NP-BSA-BMP2 groups).

## 4. Conclusions

In this work, a delivery PLGA-nanosystem previously developed for model proteins was chosen as the reference system to carry and deliver the growth factor BMP-2. This NP system, with a dual size distribution, was developed following a double-emulsion formulation in which the process and the components used were optimized to reach the appropriate colloidal and biological behavior. Encapsulation and adsorption are two different processes to load BMP-2 in PLGA NPs. Both were tested to elucidate the factors controlling them and their influence in the physico-chemical and biological properties of nanosystems. We verified that protein–polymer specific interactions have a major role in the way that protein molecules are carried and delivered from NPs. In vitro experiments showed that BMP-2-loaded PLGA NPs are the nanocarriers with the best release profile over the short-term without an initial burst and with moderate and sustained release of active protein before the onset of polymer degradation. Therefore, the biological activity is positive with no negative interaction with migration or proliferation but rather the induction of cell differentiation through the expression of ALP.

**Supplementary Materials:** The following are available online at www.mdpi.com/xxx/s1, Figure S1. Scheme of the formulation of NP-BMP2; Figure S2: Scheme of the protein adsorption process for NP-BSA-BMP2; Video S1. NTA experiments for NP-BMP2; Video S2. NTA experiments for empty NPs.

**Author Contributions:** Conceptualization, J.M.P.-G. and P.G.-M.; methodology, J.M.P.-G., A.B.J.-R. and M.P.-M.; investigation, T.d.C.-S., I.O.-O., J.M.P.-G., A.B.J.-R. and M.P.-M.; resources, A.B.J.-R., P.G.-M., F.O.-R.; writing—original draft preparation, J.M.P.-G. and M.P.-M.; writing—review and editing, J.M.P.-G., M.P.-M., A.B.J.-R., T.d.C.-S.; supervision, J.M.P.-G., P.G.-M. and F.O.-R.; funding acquisition, A.B.J.-R. and P.G.-M.

**Funding:** This research was funded by the Consejería de Economía, Innovación, Ciencia y Empleo de la Junta de Andalucía (Spain) through research groups FQM-115 and CTS-1028, by the following research project: MAT2013-43922-R—European FEDER support included—(MICINN, Spain) and by MIS Ibérica S.L.

**Acknowledgments:** The authors wish to express their appreciation for the technical support to D. Darío Abril-García.

**Conflicts of Interest:** The authors declare no conflict of interest.

**References**

1. Van Rijt, S.; Habibovic, P. Enhancing regenerative approaches with nanoparticles. *J. R. Soc. Interface* **2017**, *14*, doi:10.1098/rsif.2017.0093.
2. Kumar, B.; Jalodia, K.; Kumar, P.; Gautam, H.K. Recent advances in nanoparticle-mediated drug delivery. *J. Drug Deliv. Sci. Technol.* **2017**, *41*, 260–268, doi:10.1016/j.jddst.2017.07.019.
3. Mir, M.; Ahmed, N.; Rehman, A.U.R. Recent applications of PLGA based nanostructures in drug delivery. *Colloids Surf. B Biointerfaces* **2017**, *159*, 217–231, doi:10.1016/j.colsurfb.2017.07.038.
4. Jana, S.; Jana, S. Natural polymeric biodegradable nanoblend for macromolecules delivery. In *Recent Developments in Polymer Macro, Micro and Nano Blends*; Woodhead Publishing: Cambridge, UK, 2017; pp. 289–312, ISBN 9780081004081.
5. Danhier, F.; Ansorena, E.; Silva, J.M.; Coco, R.; Le Breton, A.; Préat, V. PLGA-based nanoparticles: An overview of biomedical applications. *J. Control. Release* **2012**, *161*, 505–522, doi:10.1016/j.jconrel.2012.01.043.
6. Ding, D.; Zhu, Q. Recent advances of PLGA micro/nanoparticles for the delivery of biomacromolecular therapeutics. *Mater. Sci. Eng. C* **2018**, *92*, 1041–1060, doi:10.1016/j.msec.2017.12.036.
7. Arias, J.L.; Unciti-Broceta, J.D.; Maceira, J.; del Castillo, T.; Hernández-Quero, J.; Magez, S.; Soriano, M.; García-Salcedo, J.A. Nanobody conjugated PLGA nanoparticles for active targeting of African Trypanosomiasis. *J. Control. Release* **2015**, *197*, 190–198, doi:10.1016/j.jconrel.2014.11.002.
8. Giteau, A.; Venier-Julienne, M.C.; Aubert-Pouëssel, A.; Benoit, J.P. How to achieve sustained and complete protein release from PLGA-based microparticles? *Int. J. Pharm.* **2008**, *350*, 14–26, doi:10.1016/j.ijpharm.2007.11.012.
9. Fredenberg, S.; Wahlgren, M.; Reslow, M.; Axelsson, A. The mechanisms of drug release in poly(lactic-co-glycolic acid)-based drug delivery systems—A review. *Int. J. Pharm.* **2011**, *415*, 34–52, doi:10.1016/j.ijpharm.2011.05.049.
10. White, L.J.; Kirby, G.T.S.; Cox, H.C.; Qodratnama, R.; Qutachi, O.; Rose, F.R.A.J.; Shakesheff, K.M. Accelerating protein release from microparticles for regenerative medicine applications. *Mater. Sci. Eng. C* **2013**, *33*, 2578–2583, doi:10.1016/j.msec.2013.02.020.
11. Ortega-Oller, I.; del Castillo-Santaella, T.; Padial-Molina, M.; Galindo-Moreno, P.; Jódar-Reyes, A.B.; Peula-García, J.M. Dual delivery nanosystem for biomolecules. Formulation, characterization, and in vitro release. *Colloids Surf. B Biointerfaces* **2017**, *159*, 586–595, doi:10.1016/j.colsurfb.2017.08.027.
12. McClements, D.J. Encapsulation, protection, and delivery of bioactive proteins and peptides using nanoparticle and microparticle systems: A review. *Adv. Colloid Interface Sci.* **2018**, *253*, 1–22, doi:10.1016/j.cis.2018.02.002.
13. Ortega-Oller, I.; Padial-Molina, M.; Galindo-Moreno, P.; O'Valle, F.; Jódar-Reyes, A.B.; Peula-García, J.M. Bone Regeneration from PLGA Micro-Nanoparticles. *BioMed Res. Int.* **2015**, *2015*, 1–18, doi:10.1155/2015/415289.
14. Bapat, R.A.; Joshi, C.P.; Bapat, P.; Chaubal, T.V.; Pandurangappa, R.; Jnanendrappa, N.; Gorain, B.; Khurana, S.; Kesharwani, P. The use of nanoparticles as biomaterials in dentistry. *Drug Discov. Today* **2019**, *24*, 85–98, doi:10.1016/j.drudis.2018.08.012.
15. Ji, Y.; Xu, G.P.; Zhang, Z.P.; Xia, J.J.; Yan, J.L.; Pan, S.H. BMP-2/PLGA delayed-release microspheres composite graft, selection of bone particulate diameters, and prevention of aseptic inflammation for bone tissue engineering. *Ann. BioMed. Eng.* **2010**, *38*, 632–639, doi:10.1007/s10439-009-9888-6.
16. Kirby, G.T.S.; White, L.J.; Rahman, C.V.; Cox, H.C.; Qutachi, O.; Rose, F.R.A.J.; Hutmacher, D.W.; Shakesheff, K.M.; Woodruff, M.A. PLGA-Based Microparticles for the Sustained Release of BMP-2. *Polymers* **2011**, *3*, 571–586, doi:10.3390/polym3010571.
17. Qutachi, O.; Shakesheff, K.M.; Buttery, L.D.K. Delivery of definable number of drug or growth factor loaded poly(dl-lactic acid-co-glycolic acid) microparticles within human embryonic stem cell derived aggregates. *J. Control. Release* **2013**, *168*, 18–27, doi:10.1016/j.jconrel.2013.02.029.
18. Wang, Y.; Wei, Y.; Zhang, X.; Xu, M.; Liu, F.; Ma, Q.; Cai, Q.; Deng, X. PLGA/PDLLA core-shell submicron spheres sequential release system: Preparation, characterization and promotion of bone regeneration in vitro and in vivo. *Chem. Eng. J.* **2015**, *273*, 490–501, doi:10.1016/j.cej.2015.03.068.

19. Zhang, H.-X.; Zhang, X.-P.; Xiao, G.-Y.; Hou, -Y.; Cheng, L.; Si, M.; Wang, S.-S.; Li, Y.-H.; Nie, L. In vitro and in vivo evaluation of calcium phosphate composite scaffolds containing BMP-VEGF loaded PLGA microspheres for the treatment of avascular necrosis of the femoral head. *Mater. Sci. Eng. C* **2016**, *60*, 298–307, doi:10.1016/j.msec.2015.11.055.
20. Begam, H.; Nandi, S.K.; Kundu, B.; Chanda, A. Strategies for delivering bone morphogenetic protein for bone healing. *Mater. Sci. Eng. C* **2017**, *70*, 856–869, doi:10.1016/j.msec.2016.09.074.
21. Balmayor, E.R.; Feichtinger, G.A.; Azevedo, H.S.; Van Griensven, M.; Reis, R.L. Starch-poly-ε-caprolactone microparticles reduce the needed amount of BMP-2. *Clin. Orthop. Relat. Res.* **2009**, *467*, 3138–3148, doi:10.1007/s11999-009-0954-z.
22. Xu, Y.; Kim, C.S.; Saylor, D.M.; Koo, D. Polymer degradation and drug delivery in PLGA-based drug–polymer applications: A review of experiments and theories. *J. BioMed. Mater. Res. Part B Appl. Biomater.* **2017**, *105*, 1692–1716, doi:10.1002/jbm.b.33648.
23. Padial-Molina, M.; de Buitrago, J.G.; Sainz-Urruela, R.; Abril-Garcia, D.; Anderson, P.; O'Valle, F.; Galindo-Moreno, P. Expression of Musashi-1 during osteogenic differentiation of oral MSC: An in vitro study. *Int. J. Mol. Sci.* **2019**, *20*, 2171, doi:10.3390/ijms20092171.
24. D'Angelo, I.; Garcia-Fuentes, M.; Parajó, Y.; Welle, A.; Vántus, T.; Horváth, A.; Bökönyi, G.; Kéri, G.; Alonso, M.J. Nanoparticles based on PLGA:poloxamer blends for the delivery of proangiogenic growth factors. *Mol. Pharm.* **2010**, *7*, 1724–1733, doi:10.1021/mp1001262.
25. Chang, H.-C.; Yang, C.; Feng, F.; Lin, F.-H.; Wang, C.-H.; Chang, P.-C. Bone morphogenetic protein-2 loaded poly(D,L-lactide-co-glycolide) microspheres enhance osteogenic potential of gelatin/hydroxyapatite/β-tricalcium phosphate cryogel composite for alveolar ridge augmentation. *J. Formos. Med. Assoc.* **2017**, *116*, 973–981, doi:10.1016/j.jfma.2017.01.005.
26. Padial-Molina, M.; Volk, S.L.; Rios, H.F. Periostin increases migration and proliferation of human periodontal ligament fibroblasts challenged by tumor necrosis factor -α and Porphyromonas gingivalis lipopolysaccharides. *J. Periodontal Res.* **2014**, *49*, 405–414, doi:10.1111/jre.12120.
27. Liang, C.-C.; Park, A.Y.; Guan, J.-L. In vitro scratch assay: A convenient and inexpensive method for analysis of cell migration in vitro. *Nat. Protoc.* **2007**, *2*, 329–333, doi:10.1038/nprot.2007.30.
28. Houghton, P.; Fang, R.; Techatanawat, I.; Steventon, G.; Hylands, P.J.; Lee, C.C. The sulphorhodamine (SRB) assay and other approaches to testing plant extracts and derived compounds for activities related to reputed anticancer activity. *Methods* **2007**, *42*, 377–387, doi:10.1016/j.ymeth.2007.01.003.
29. Iqbal, M.; Zafar, N.; Fessi, H.; Elaissari, A. Double emulsion solvent evaporation techniques used for drug encapsulation. *Int. J. Pharm.* **2015**, *496*, 173–190, doi:10.1016/j.ijpharm.2015.10.057.
30. Sánchez-Moreno, P.; Ortega-Vinuesa, J.L.; Boulaiz, H.; Marchal, J.A.; Peula-García, J.M. Synthesis and characterization of lipid immuno-nanocapsules for directed drug delivery: Selective antitumor activity against HER2 positive breast-cancer cells. *Biomacromolecules* **2013**, *14*, 4248–4259, doi:10.1021/bm401103t.
31. Lochmann, A.; Nitzsche, H.; von Einem, S.; Schwarz, E.; Mäder, K. The influence of covalently linked and free polyethylene glycol on the structural and release properties of rhBMP-2 loaded microspheres. *J. Control. Release* **2010**, *147*, 92–100, doi:10.1016/j.jconrel.2010.06.021.
32. Kempen, D.H.R.; Lu, L.; Hefferan, T.E.; Creemers, L.B.; Maran, A.; Classic, K.L.; Dhert, W.J.A.; Yaszemski, M.J. Retention of in vitro and in vivo BMP-2 bioactivities in sustained delivery vehicles for bone tissue engineering. *Biomaterials* **2008**, *29*, 3245–3252, doi:10.1016/j.biomaterials.2008.04.031.
33. Santander-Ortega, M.J.; Csaba, N.; González, L.; Bastos-González, D.; Ortega-Vinuesa, J.L.; Alonso, M.J. Protein-loaded PLGA–PEO blend nanoparticles: Encapsulation, release and degradation characteristics. *Colloid Polym. Sci.* **2010**, *288*, 141–150, doi:10.1007/s00396-009-2131-z.
34. Chung, Y.I.; Ahn, K.M.; Jeon, S.H.; Lee, S.Y.; Lee, J.H.; Tae, G. Enhanced bone regeneration with BMP-2 loaded functional nanoparticle-hydrogel complex. *J. Control. Release* **2007**, *121*, 91–99, doi:10.1016/j.jconrel.2007.05.029.
35. La, W.-G.; Kang, S.-W.; Yang, H.S.; Bhang, S.H.; Lee, S.H.; Park, J.-H.; Kim, B.-S. The Efficacy of Bone Morphogenetic Protein-2 Depends on Its Mode of Delivery. *Artif. Organs* **2010**, *34*, 1150–1153, doi:10.1111/j.1525-1594.2009.00988.x.
36. Fu, Y.; Du, L.; Wang, Q.; Liao, W.; Jin, Y.; Dong, A.; Chen, C.; Li, Z. In vitro sustained release of recombinant human bone morphogenetic protein-2 microspheres embedded in thermosensitive hydrogels. *Die Pharm.* **2012**, *67*, 299–303, doi:10.1691/ph.2012.1649.

37. Rahman, C.V.; Ben-David, D.; Dhillon, A.; Kuhn, G.; Gould, T.W.A.; Müller, R.; Rose, F.R.A.J.; Shakesheff, K.M.; Livne, E. Controlled release of BMP-2 from a sintered polymer scaffold enhances bone repair in a mouse calvarial defect model. *J. Tissue Eng. Regen. Med.* **2014**, *8*, 59–66, doi:10.1002/term.1497.
38. Pakulska, M.M.; Elliott Donaghue, I.; Obermeyer, J.M.; Tuladhar, a.; McLaughlin, C.K.; Shendruk, T.N.; Shoichet, M.S. Encapsulation-free controlled release: Electrostatic adsorption eliminates the need for protein encapsulation in PLGA nanoparticles. *Sci. Adv.* **2016**, *2*, e1600519, doi:10.1126/sciadv.1600519.
39. Fu, C.; Yang, X.; Tan, S.; Song, L. Enhancing Cell Proliferation and Osteogenic Differentiation of MC3T3-E1 Pre-osteoblasts by BMP-2 Delivery in Graphene Oxide-Incorporated PLGA/HA Biodegradable Microcarriers. *Sci. Rep.* **2017**, *7*, 12549, doi:10.1038/s41598-017-12935-x.
40. Peula, J.M.; de las Nieves, F.J. Adsorption of monomeric bovine serum albumin on sulfonated polystyrene model colloids 1. Adsorption isotherms and effect of the surface charge density. *Colloids Surf. A Physicochem. Eng. Asp.* **1993**, *77*, 199–208, doi:10.1016/0927-7757(93)80117-W.
41. Peula, J.M.; de las Nieves, F.J. Adsorption of monomeric bovine serum albumin on sulfonated polystyrene model colloids 3. Colloidal stability of latex—Protein complexes. *Colloids Surf. A Physicochem. Eng. Asp.* **1994**, *90*, 55–62, doi:10.1016/0927-7757(94)02889-3.
42. Peula, J.M.; Hidalgo-Alvarez, R.; De Las Nieves, F.J. Coadsorption of IgG and BSA onto sulfonated polystyrene latex: I. Sequential and competitive coadsorption isotherms. *J. Biomater. Sci. Polym. Ed.* **1996**, *7*, 231–240, doi:10.1163/156856295X00274.
43. Siafaka, P.I.; Üstündağ Okur, N.; Karavas, E.; Bikiaris, D.N. Surface modified multifunctional and stimuli responsive nanoparticles for drug targeting: Current status and uses. *Int. J. Mol. Sci.* **2016**, *17*, 1440, doi:10.3390/ijms17091440.
44. Peula-García, J.M.; Hidalgo-Alvarez, R.; De Las Nieves, F.J. Colloid stability and electrokinetic characterization of polymer colloids prepared by different methods. *Colloids Surf. A Physicochem. Eng. Asp.* **1997**, *127*, 19–24, doi:10.1016/S0927-7757(96)03890-3.
45. Santander-Ortega, M.J.; Lozano-López, M.V.; Bastos-González, D.; Peula-García, J.M.; Ortega-Vinuesa, J.L. Novel core-shell lipid-chitosan and lipid-poloxamer nanocapsules: Stability by hydration forces. *Colloid Polym. Sci.* **2010**, *288*, 159–172, doi:10.1007/s00396-009-2132-y.
46. Peula-Garcia, J.M.; Hidaldo-Alvarez, R.; De las Nieves, F.J. Protein co-adsorption on different polystyrene latexes: Electrokinetic characterization and colloidal stability. *Colloid Polym. Sci.* **1997**, *275*, 198–202, doi:10.1007/s003960050072.
47. Santander-Ortega, M.J.; Bastos-González, D.; Ortega-Vinuesa, J.L. Electrophoretic mobility and colloidal stability of PLGA particles coated with IgG. *Colloids Surf. B Biointerfaces* **2007**, *60*, 80–88, doi:10.1016/j.colsurfb.2007.06.002.
48. Peula, J.M.; Callejas, J.; de las NIeves, F.J. Adsorption of Monomeric Bovine Serum Albumin on Sulfonated Polystyrene Model Colloids. II. Electrokinetic Characterization of Latex-Protein Complexes. In *Surface Properties of Biomaterials*; Butterworth and Heinemann: Oxford, UK, 1994; pp. 61–69.
49. Sun, D. Effect of Zeta Potential and Particle Size on the Stability of SiO2 Nanospheres as Carrier for Ultrasound Imaging Contrast Agents. *Int. J. Electrochem. Sci.* **2016**, 8520–8529, doi:10.20964/2016.10.30.
50. del Castillo-Santaella, T.; Peula-García, J.M.; Maldonado-Valderrama, J.; Jódar-Reyes, A.B. Interaction of surfactant and protein at the O/W interface and its effect on colloidal and biological properties of polymeric nanocarriers. *Colloids Surf. B Biointerfaces* **2019**, *173*, 295–302, doi:10.1016/j.colsurfb.2018.09.072.
51. Schrier, J.A.; DeLuca, P.P. Porous bone morphogenetic protein-2 microspheres: Polymer binding and in vitro release. *AAPS PharmSciTech* **2001**, *2*, 66–72, doi:10.1208/pt020317.
52. Padial-Molina, M.; O'Valle, F.; Lanis, A.; Mesa, F.; Dohan Ehrenfest, D.M.; Wang, H.-L.; Galindo-Moreno, P. Clinical application of mesenchymal stem cells and novel supportive therapies for oral bone regeneration. *BioMed Res. Int.* **2015**, *2015*, doi:10.1155/2015/341327.
53. Inai, K.; Norris, R.A.; Hoffman, S.; Markwald, R.R.; Sugi, Y. BMP-2 induces cell migration and periostin expression during atrioventricular valvulogenesis. *Dev. Biol.* **2008**, *315*, 383–396, doi:10.1016/j.ydbio.2007.12.028.
54. Gamell, C.; Osses, N.; Bartrons, R.; Rückle, T.; Camps, M.; Rosa, J.L.; Ventura, F.; Imamura, T. BMP2 induction of actin cytoskeleton reorganization and cell migration requires PI3-kinase and Cdc42 activity. *J. Cell Sci.* **2008**, *121*, 3960–3970, doi:10.1242/jcs.031286.

55. Friedrichs, M.; Wirsdöerfer, F.; Flohé, S.B.; Schneider, S.; Wuelling, M.; Vortkamp, A. BMP signaling balances proliferation and differentiation of muscle satellite cell descendants. *BMC Cell Biol.* **2011**, *12*, 26, doi:10.1186/1471-2121-12-26.
56. Hrubi, E.; Imre, L.; Robaszkiewicz, A.; Virág, L.; Kerényi, F.; Nagy, K.; Varga, G.; Jenei, A.; Hegedüs, C. Diverse effect of BMP-2 homodimer on mesenchymal progenitors of different origin. *Hum. Cell* **2018**, *31*, 139–148, doi:10.1007/s13577-018-0202-5.
57. Kim, H.K.W.; Oxendine, I.; Kamiya, N. High-concentration of BMP2 reduces cell proliferation and increases apoptosis via DKK1 and SOST in human primary periosteal cells. *Bone* **2013**, *54*, 141–150, doi:10.1016/j.bone.2013.01.031.